\documentclass{iau}
\usepackage{graphicx,natbib}
 
\newcommand{\apj}{ApJ}           
\newcommand{\mnras}{MNRAS}       

\newcommand{\aap}{A\&A}

\newcommand{\aj}{AJ}

\title{A kinematic analysis of the Giant star-forming Region of N11}

\author[Torres-Flores et al.]{Sergio Torres-Flores$^1$, Rodolfo Barb\'a$^1$, Jes\'us Ma\'iz Apell\'aniz$^2$, M\'onica Rubio$^3$ \and Guillermo Bosch$^4$}

\affiliation{$^1$Departamento de F\'isica, Universidad de La Serena, \\Av. Cisternas 1200 Norte, La Serena, Chile \\
email: {\tt storres@dfuls.cl} \\
$^2$ Centro de Astrobiolog\'ia (INTA-CSIC), ESAC campus, P.O. Box 78,\\ 28691 Villanueva de la Ca\~nada, Madrid, Spain\\
$^3$ Departamento de Astronom\'ia, Universidad de Chile, Casilla 36-D, Santiago, Chile\\
$^4$ Facultad de Ciencias Astron\'omicas y Geof\'isicas, Universidad Nacional de la La Plata,\\ Paseo del Bosque s/n, 1900 La Plata, Argentina  
}

\pubyear{2014}
\volume{309}
\jname{Galaxies in 3D across the Universe}
\editors{B. L. Ziegler, F. Combes, H. Dannerbauer, M. Verdugo, eds.}

\begin{document}

\maketitle

\begin{abstract}

In this work we present high resolution spectroscopic data of the giant star-forming region of N11, obtained with the GIRAFFE instrument at the Very Large Telescope. By using this data set, we find that most of the H$\alpha$ emission lines profiles in this complex can be fitted by a single Gaussian, however, multiple emission line profiles can be observed in the central region of N11. By adding all the spectra, we derive the integrated H$\alpha$ profile of this complex, which displays a width ($\sigma$) of about 12 km s$^{-1}$ (corrected by instrumental and thermal width). We find that a single Gaussian fit on the integrated H$\alpha$ profile leaves remaining wings, which can be fitted by a secondary broad Gaussian component. In addition, we find high velocity features, which spatially correlate with soft diffuse X-ray emission.
\end{abstract}

\firstsection
\section{Introduction and Data}

Giant HII regions (GHIIR) are characterized by displaying integrated emission lines which have supersonic widths (\citealt{smith72}, \citealt{melnick99}), however, the origin of these motions is still unclear. In addition, some authors have reported the existence of a low-intensity broad component in the integrated emission line profiles of GHIIR. Some authors suggested that this broad component is produced by a highly turbulent diffuse component \cite{melnick99}, while others authors argue that this component arises as a superposition of multiple bubbles in expansion \cite{chu94}. In order to study these phenomena, we present a kinematic analysis of the ionized gas in the star-forming region N11 (see previous works in \citealt{rosado96}, \citealt{naze04}). We have used spectroscopic data taken with the GIRAFFE instrument (MEDUSA mode), at the Very Large Telescope to derive a data cube with a size of 19'$\times$16', where the MEDUSA fibers were located in a quasi-regular grid over N11. In the following we describe part of our results, which are based on the kinematics of the H$\alpha$ emission line.

\section{Summary of the main results}

\begin{figure}
\centering
\includegraphics[width=0.42\columnwidth]{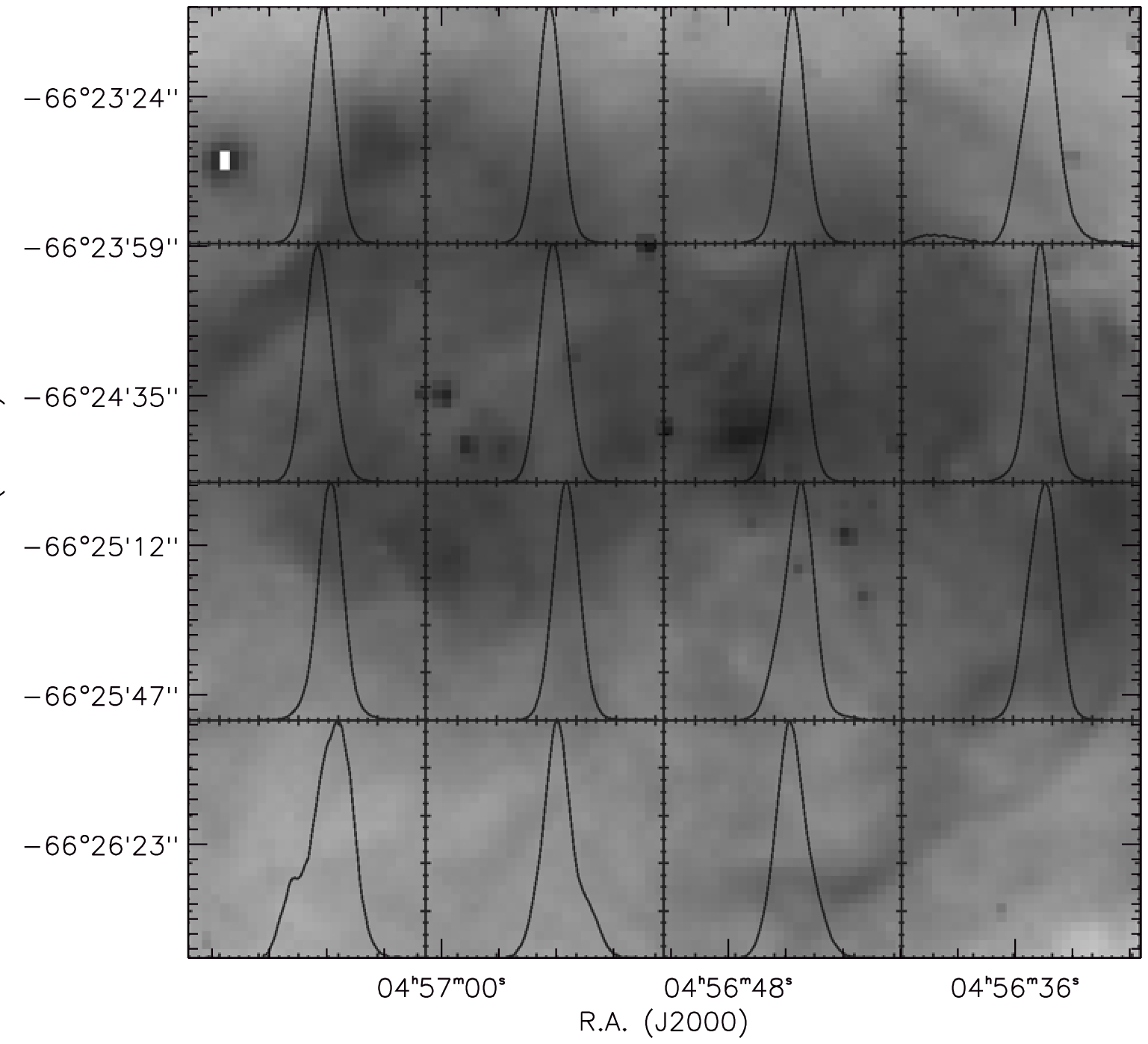} 
\includegraphics[width=0.42\columnwidth]{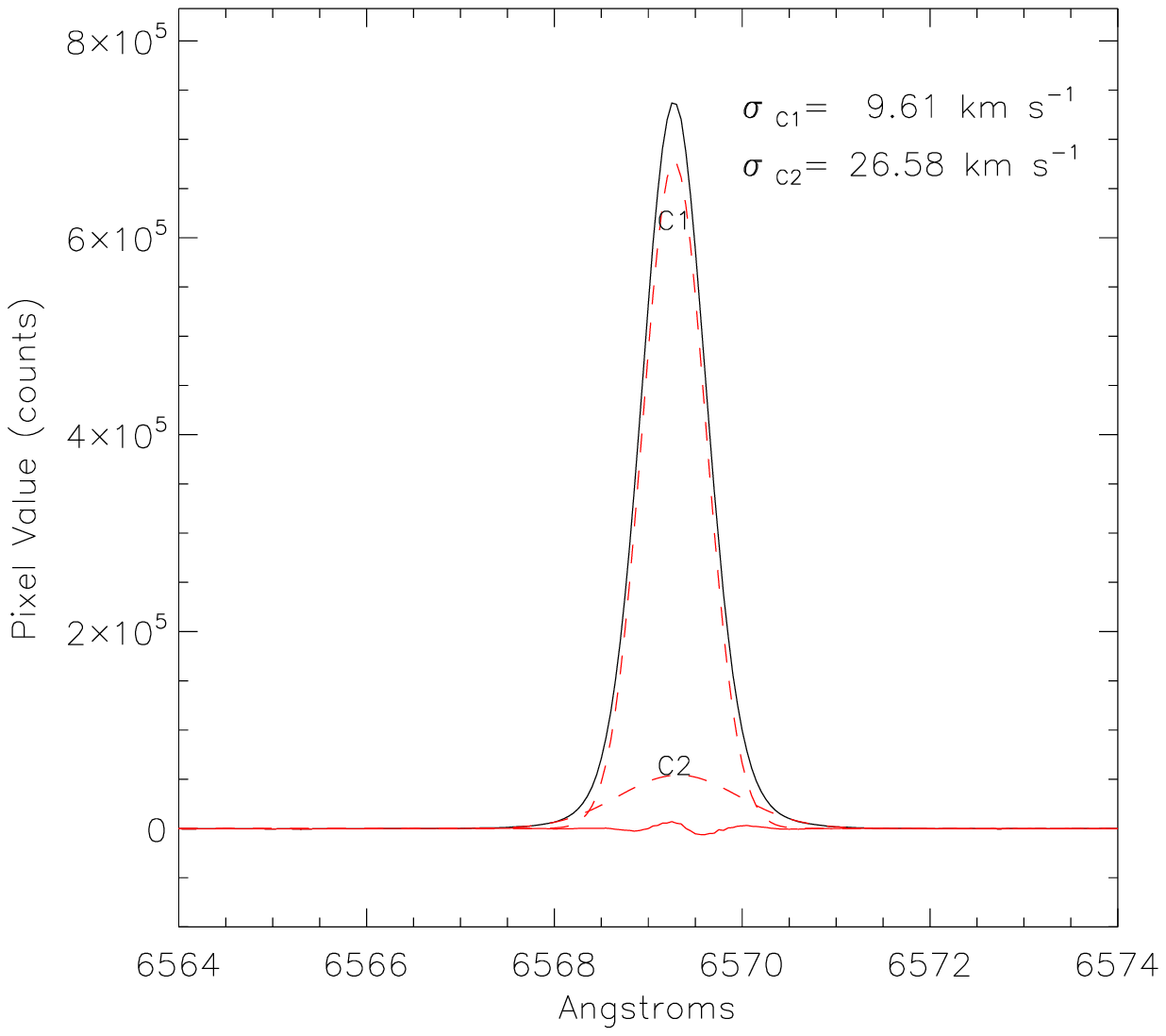} 
\caption{\textit{Left panel:} An example of our spectroscopic H$\alpha$ data cube, where the profiles are superimposed on a H$\alpha$ image of N11B. \textit{Right panel}: Integrated H$\alpha$ profile of N11, derived from the whole H$\alpha$ data cube (black line). Dashed red lines represent the Gaussian fits, and the solid red line shows the residual between the Gaussian fits and the observations.}
\label{fig1}
\end{figure}

In the left panel of Figure \ref{fig1} we plot an example of the H$\alpha$ data cube over an H$\alpha$ image of N11B, which is a star-forming complex in N11 (\citealt{barba03}). On this image (where the north is up and east is to the left), we can see the H$\alpha$ emission line, which displays slightly asymmetric profiles, despite the fact that we are observing an actively star-forming system. At the south of N11B we can observe asymmetric H$\alpha$ profiles, which are produced by high velocity components and/or small expanding structures. We note that we have found high velocity features close to the central cavity in N11, where soft X-ray emission has been reported. 

In the right panel of Figure \ref{fig1} we show the integrated H$\alpha$ profile of N11 (solid black line), which has been derived from the entire GIRAFFE/MEDUSA data cube (19'$\times$16'). In order to determine the width of this profile, we have fitted a single Gaussian on it, obtaining a width of \mbox{$\sigma \sim$ 12 km s$^{-1}$} (once corrected by instrumental and thermal widths). However, we detected the presence of remaining wings in the residual emission. For this reason, we have fitted a secondary Gaussian component to remove these wings. This exercise produce a narrow high-intensity component and a broad low-intensity component, which are plotted in Figure \ref{fig1} with red dashed lines. On the same figure we plot the difference between the observed integrated profile and the Gaussian fits, which is negligible (red solid line). We found that the narrow component has a corrected width of \mbox{$\sigma \sim$ 10 km s$^{-1}$}. This width, which may arises from the contribution of the stellar feedback, supernova events and gravity, is lower than the observed value for the 30 Dor nebula (\citealt{torresflores13}), suggesting that the evolutionary stage of the nebula may play an important role in the width of the integrated H$\alpha$ profile.

\section*{Acknowledgements}

\noindent

We would like to thank the conference organizers. ST-F acknowledges the financial support of the project CONICYT PAI/ACADEMIA 7912010004.

\end{document}